\documentclass[a4paper,11pt]{article}

\usepackage[UKenglish]{babel}

\usepackage{jheppub}
\usepackage{physics}
\usepackage{amssymb}
\usepackage{mathrsfs}
\usepackage{kbordermatrix}
\usepackage[dvipsnames]{xcolor}
\usepackage{graphicx}
\usepackage{cleveref}
\usepackage{csquotes}
\usepackage{xparse}
\usepackage[backend=biber,style=numeric-comp,sorting=none,backref,maxnames=10]{biblatex}

\usepackage{etoolbox}
\patchcmd{\abstract}{\small}{}{}{}

\usepackage{tikz}
\usetikzlibrary{3d}
\usetikzlibrary{calc}
\usetikzlibrary{decorations.pathreplacing}
\usetikzlibrary{decorations.pathmorphing}
\usetikzlibrary{decorations.markings}
\usetikzlibrary{arrows.meta}
\usetikzlibrary{cd}

\tikzset{->-/.style={decoration={
  markings,
  mark=at position 0.5 with {\arrow{Latex}}},postaction={decorate}}}

\makeatletter
\def\tikz@auto@anchor{%
    \pgfmathtruncatemacro\angle{atan2(\pgf@y,\pgf@x)-90}
    \edef\tikz@anchor{\angle}%
}
\def\tikz@auto@anchor@prime{%
    \pgfmathtruncatemacro\angle{atan2(\pgf@y,\pgf@x)+90}
    \edef\tikz@anchor{\angle}%
}
\makeatother

\tikzset{dotnode/.style={inner sep=0pt,outer sep=0pt,
  circle,fill,minimum size=5pt}}

\newcommand{\nodename}[2]{\pref x#1x#2}
\NewDocumentCommand{\drawTriangle}{mmmmo}{
  \drawNodes{#1}{#2}{#3}{#4}[#5]
  \drawArrows{#1}{#2}{#3}{#4}[#5]
}
\NewDocumentCommand{\drawNodes}{mmmmo}{
  \def\n{#1}
  \def\nA{#2}
  \def\nB{#3}
  \def\nC{#4}
  \def\pref{#5}
  \pgfmathsetmacro{\npp}{\n+1}
  \coordinate (BmA) at ($(\nB)-(\nA)$);
  \coordinate (CmA) at ($(\nC)-(\nA)$);
  \draw (\nA) -- (\nB) -- (\nC) -- (\nA);
  \begin{scope}[orange,nodes=dotnode]
    \foreach \i in {0,...,\npp}{
      \pgfmathparse{\npp-\i}
      \foreach \j in {0,...,\pgfmathresult} {
        \pgfmathparse{!(\i == 0 && \j == 0)
                      && !(\i == \npp && \j == 0)
                      && !(\i == 0 && \j == \npp)}
        \ifnum\pgfmathresult=1
          \node (\nodename{\i}{\j}) at
            ($(\nA)+{\i/\npp}*(BmA)+{\j/\npp}*(CmA)$)
            {};
        \fi
      }
    }
  \end{scope}
}

\NewDocumentCommand{\drawArrows}{mmmmo}{
  \def\n{#1}
  \def\nA{#2}
  \def\nB{#3}
  \def\nC{#4}
  \def\pref{#5}
  \pgfmathsetmacro{\npp}{\n+1}
  \begin{scope}[orange,-latex]
    \foreach \i in {1,...,\n}{
      \pgfmathparse{\n-\i}
      \foreach \j in {0,...,\pgfmathresult} {
        \pgfmathtruncatemacro{\newJ}{\j+1}
        \draw (\nodename{\i}{\j}) -- (\nodename{\i}{\newJ});
      }
    }
    \foreach \i in {1,...,\n}{
      \pgfmathparse{\npp-\i}
      \foreach \j in {1,...,\pgfmathresult} {
        \pgfmathtruncatemacro{\newI}{\i-1}
        \draw (\nodename{\i}{\j}) -- (\nodename{\newI}{\j});
      }
    }
    \pgfmathparse{\n-1}
    \foreach \i in {0,...,\pgfmathresult}{
      \pgfmathparse{\n-\i}
      \foreach \j in {1,...,\pgfmathresult} {
        \pgfmathtruncatemacro{\newI}{\i+1}
        \pgfmathtruncatemacro{\newJ}{\j-1}
        \draw (\nodename{\i}{\j}) -- (\nodename{\newI}{\newJ});
      }
    }
  \end{scope}
}

\newcommand{\ii}{\mathrm{i}}
\newcommand{\e}{\mathrm{e}}
\newcommand{\alphae}{\alpha_\mathrm{e}}
\newcommand{\alpham}{\alpha_\mathrm{m}}
\newcommand{\qe}{q_\mathrm{e}}
\newcommand{\qm}{q_\mathrm{m}}
\newcommand{\N}{\mathbb{N}}
\newcommand{\Z}{\mathbb{Z}}

\newcommand{\R}{\mathbb{R}}
\newcommand{\QQ}{\mathcal{Q}}
\newcommand{\cS}{\mathcal{S}}
\newcommand{\sN}{\mathcal{N}}
\newcommand{\dg}{\mathbb{D}^{(1)}}
\newcommand{\M}{\mathcal{M}}
\newcommand{\longto}{\longrightarrow}
\newcommand{\xhookrightarrow}[1]{\lhook\joinrel\xrightarrow{#1}}
\DeclareMathOperator{\coker}{coker}
\DeclareMathOperator{\im}{im}
\DeclareMathOperator{\Hom}{Hom}
\DeclareMathOperator{\Tor}{Tor}
\DeclareMathOperator{\Ext}{Ext}
\DeclareMathOperator{\SU}{SU}
\DeclareMathOperator{\U}{U}
\DeclareMathOperator{\g}{\mathfrak{g}}
\DeclareMathOperator{\su}{\mathfrak{su}}
\DeclareMathOperator{\hodge}{\star}

\pdfoutput=1

\title{Defect groups of class $\cS$ theories from the Coulomb branch}
\author{Elias Riedel Gårding}
\affiliation{Department of Mathematics, Uppsala University\\
  Box 480, SE-75106 Uppsala, Sweden}
\emailAdd{elias.riedel\_garding@math.uu.se}

\abstract{%
  We study the global forms of class $\cS[A_{N-1}]$ 4d $\sN = 2$ theories by
  deriving their defect groups (charges of line operators up to screening by
  local operators) from Coulomb branch data. Specifically, we employ an explicit
  construction of the BPS quiver for the case of full regular punctures to
  show that the defect group is $(\Z_N)^{2g}$, where $g$ is the genus of the
  associated Riemann surface. This determines a sector of surface operators in
  the 5d symmetry TFT. We show how these can also be identified from dimensional
  reduction of M-theory. We discuss connections to the theory of cluster
  algebras.}

\begin{document}
\maketitle
\flushbottom

\section{Introduction and overview}
In exploring properties of quantum field theories that cannot be accessed
through perturbative methods, symmetry is one of the precious few footholds
available. In particular, anomalies of global symmetries provide quantities
that, by the classic anomaly matching argument of \textcite{tHooft:1979rat},
are invariant under renormalisation group flow.

A modern viewpoint on anomalies is in terms of invertible theories in one
dimension higher, so that an anomalous theory lives on the boundary of an
anomaly theory. Upon a background gauge transformation, the anomalous phases
from the boundary and the bulk cancel, rendering the combined system
anomaly-free; this is \emph{anomaly inflow}
\cite{Callan:1984sa,Freed:2014iua,Freed:2023snr}. In this framework, anomaly
matching is the statement that anomaly theories are topological and therefore
invariant under RG flow.

The concept of a symmetry, traditionally seen as a transformation on the fields
that leaves the action and (up to an anomalous phase) partition function
invariant, has in recent years been reexamined and generalised. The key
observation is that the presence of a traditional symmetry is equivalent to the
existence of operators/defects, supported on closed codimension-1 submanifolds
of spacetime, and invariant under continuous deformations of those submanifolds.
The group law of the symmetry is expressed in the fusion algebra of the
corresponding defects. This perspective suggests the generalisation to $p$-form
symmetries \cite{Kapustin:2014gua,Gaiotto:2014kfa}, whose charged objects are
$p$-dimensional extended operators acted on by codimension-$(p + 1)$ symmetry
operators. These may in addition mix with each other in higher group-like
structures \cite{Cordova:2018cvg}. Another natural generalisation is to
non-invertible symmetries
\cite{Bhardwaj:2022yxj,Bashmakov:2022jtl,Bashmakov:2022uek,
  Kaidi:2021xfk,Kaidi:2022cpf,Choi:2022fgx,Choi:2022jqy}.

A similar picture to that of anomaly inflow applies when one has several field
theories with identical local dynamics, but different spectra of extended
operators, that is, different global structures
\cite{Gaiotto:2010be,Aharony:2013hda}. The theories can then be viewed together
as a \emph{relative} field theory \cite{Freed:2012bs,Bhardwaj:2021mzl} living on
the boundary of a \emph{non-invertible} TQFT, called the \emph{symmetry TFT}
(SymTFT) \cite{Kapustin:2014gua,Gaiotto:2020iye,Freed:2022qnc,Apruzzi:2021nmk},
in one dimension higher. The set of all global forms is a property of the SymTFT
itself, while a topological boundary condition picks out a boundary theory with
a particular global structure (see \cref{fig:sandwich} below). Indeed, the
SymTFT and its topological boundary conditions can be studied abstractly, quite
apart from any dynamical boundary theory, just as a group can be studied
independently of its representations \cite{Freed:2022qnc}.

In the remainder of this section, we outline the structure of the paper,
introduce the central concepts, summarise the main points and give some
directions for future work.

\Textcite{DelZotto:2022ras} discussed a concrete example of the relation
between global structures and the symmetry TFT, namely the case of a 4d QFT that
has a Coulomb phase. We review it in some detail in \cref{sec:review}.

In this work, we apply the story of \cite{DelZotto:2022ras} to the case of the
class $\cS$ construction \cite{Gaiotto:2009we,Gaiotto:2009hg}. We start from the
6d $\sN = (2,0)$ SCFT of type $\g$---a relative theory---and get a 4d $\sN = 2$
supersymmetric relative theory by compactifying on a Riemann surface
$\Sigma_{g,p}$ of genus $g$ and with $p$ punctures. We study the most
straightforward case, where the punctures are all \emph{regular} and \emph{full}
(see \cite[chap.~12]{Tachikawa:2013kta} for an elaboration on this) and $\g =
A_{N - 1} = \su(N)$. We call the 4d relative theory $\cS[\su(N), \Sigma_{g,p}]$.
Concretely, we claim that the \emph{defect group} \cite{DelZotto:2015isa} (the
group $\dg$ of 1-form symmetry charges of lines that appear in some global form)
of this theory is
\begin{equation}
  \label{eq:defect-group-claim}
  \dg(\cS[\su(N), \Sigma_{g,p}]) \cong (\Z_N)^{2g}.
\end{equation}
The defect group corresponds to the group of surface operators in the 5d SymTFT;
it therefore allows us to determine the sector of the SymTFT that couples to the
1-form symmetry, as reviewed in \cref{sec:review}.

There are a priori good reasons to expect this defect group, and indeed special
cases of the claim have appeared in earlier literature.
\Textcite{Tachikawa:2013hya} considered the case without punctures and (though
the term \emph{defect group} was not established) argued that
\begin{equation}
  \label{eq:tachikawa-defect-group}
  \dg(\cS[\g, \Sigma_{g,0}]) \cong H^1(\Sigma_{g,0}; Z(G))
\end{equation}
where $Z(G)$ is the centre of the simply connected group $G$ with algebra $\g$;
this agrees with \eqref{eq:defect-group-claim}. \Textcite{Bhardwaj:2021pfz}
proposed that regular untwisted punctures do not affect the defect group in
general, so that in particular \eqref{eq:tachikawa-defect-group} holds also for
$p > 0$. They verified this expectation in the cases $\cS[\su(N),
\Sigma_{0,4}]$, $\cS[\su(2), \Sigma_{g,p}]$ and $\cS[\su(N),
\Sigma_{1,\tilde{p}}]$, where $\Sigma_{1,\tilde{p}}$ denotes the torus with $p$
\emph{simple} punctures (see \cref{sec:non-full}), as well as many theories
outside the scope of this paper. Our confirmation of
\eqref{eq:defect-group-claim} adds another class of examples to this list.

When $\g = \su(N)$ and there are no punctures, one can derive the defect group
via the SymTFT that descends from the M-theory Chern--Simons term when realising
the 6d theory as the worldvolume theory of a stack of M5-branes; we review this
in \cref{sec:m-theory} following \cite{Witten:1998wy} and
\cite{Bashmakov:2022uek}. Furthermore, we add punctures using a geometric
construction \cite{Bah:2018jrv,Bah:2019jts} and argue that the defect
group is unaffected as expected.

The main result of this work is presented in \cref{sec:bps-quivers}. We make use
of the known charge lattices of class $\cS$ theories to verify
\eqref{eq:defect-group-claim} for small $N$, $g$ and $p$ by explicitly
calculating the BPS quiver and defect group. This serves as a check of the
paradigm from \cite{DelZotto:2022ras} that the SymTFT can be accessed from
Coulomb branch data. A similar analysis for 5d and 6d theories was carried out
in \cite{Apruzzi:2022dlm}. In \cref{sec:non-full}, we do the same calculation
for a class of theories where the punctures are not full. In
\cref{sec:structure}, we make some observations on the structure of the BPS
quivers to motivate why \eqref{eq:defect-group-claim} should hold.

There are several interesting avenues for generalisation and application of our
methods in future work. Ideally, one would like a systematic, algorithmic
construction of BPS quivers for class $\cS$ theories with arbitrary simply laced
gauge algebra $\g$ and any configuration of regular punctures. Several proposals
in this direction have been made \cite{Xie:2012dw,Gabella:2017hpz}, but the
constructions are complex and it is not clear to what extent they can be
generalised. Here the work of \textcite{Goncharov:2019gzh} provides a promising
starting point. Once one knows the BPS quiver of a theory, the natural
application is to compute its BPS spectrum using the mutation method of
\cite{Alim:2011kw}. In particular, this enables one to calculate the Schur index
according to the conjecture of \textcite{Cordova:2015nma}; comparing this to the
derivation of the index as a TQFT correlator on $\Sigma_{g,p}$
\cite{Gadde:2009kb,Rastelli:2014jja} would be a good cross-check. The argument
in \cref{sec:m-theory} on adding regular punctures is somewhat schematic; it
would be useful to reproduce it at a higher level of rigour. Finally, filling in
the proof sketch of \cref{sec:structure} would establish a result of general
interest in the mathematics of cluster algebras.

\Cref{sec:computation} describes the details of the computer calculation using
BPS quivers that confirms \eqref{eq:defect-group-claim}. \Cref{sec:math}
reviews some of the standard mathematical theorems used throughout the paper.

\section{Defect group and SymTFT in the Coulomb phase}
\label{sec:review}

\subsection{Wilson and 't Hooft lines in Maxwell theory}
As preparation for the general case, let us consider four-dimensional Maxwell
theory with a $\U(1)$ $1$-form gauge field $A$. The field strength $F = \dd{A}$
is a closed but not necessarily exact $2$-form. We wish the Wilson loop
$W_{\alphae}(\gamma) = \e^{\ii \alphae \oint_\gamma A}$ of electric charge
$\alphae$ to be well-defined. Consider deforming the loop from $\gamma$ to
$\gamma'$ along a surface $\Sigma$ with $\partial\Sigma = \gamma' - \gamma$; we
find $W_{\alphae}(\gamma') = \e^{\ii\alphae \int_\Sigma F} W_{\alphae}(\gamma)$.
In particular, from $\gamma' = \gamma$ we must have $\alphae \oint_\Sigma F \in
2\pi\Z$ for all closed surfaces $\Sigma$. Next, we insert an 't Hooft loop
$H_{\alpham}(\ell)$, defined by sourcing a magnetic flux: $\oint_\Sigma
\frac{F}{2\pi} = \alpham$ whenever $\Sigma$ links $\ell$. In the presence of
$W_{\alphae}$, this is consistent with the previous condition when
$\alphae\alpham \in \Z$; this is the Dirac quantisation condition. Thus the set
of allowed Wilson lines constrain the allowed 't Hooft lines, and vice versa.

Let us now add massive dynamical states to the theory: a particle with electric
charge $\qe$ and a monopole with charge $\qm$. Integrating them out is
equivalent to inserting Wilson and 't Hooft operators $W_{\qe}$ and $H_{\qm}$
supported on their worldlines in the IR path integral; thus these must be
allowed line operators. In particular we have
\begin{equation}
  \label{eq:maxwell-defect-group}
  \qe\qm \in \Z, \quad \qe\alpham \in \Z, \quad \alphae\qm \in \Z,
\end{equation}
where $\alphae$ and $\alpham$ are the charges of any other allowed lines in the
theory. Given the dynamical charges $\qe$ and $\qm$, the set of solutions
$(\alphae \bmod{\qe}, \alpham \bmod{\qm})$ to \eqref{eq:maxwell-defect-group} up
to screening by the dynamical states is called the \emph{defect group}, in this
case $\Z_{\qe\qm} \times \Z_{\qe\qm}$. Here the most general line is dyonic,
with simultaneous electric and magnetic charges $(\alphae, \alpham)$. The
condition for two such lines to be well-defined in each others' presence is the
Dirac--Schwinger--Zwanziger quantisation condition
\cite{Schwinger:1969ib,Zwanziger1965} $\alphae\alpham' - \alpham\alphae' \in
\Z$. A maximal subgroup of lines satisfying this will be (the Pontryagin dual
of) the \emph{1-form symmetry}. We will elaborate on this in the next section.

The reasoning so far holds regardless of the normalisation of the gauge field
$A$. If the gauge transformations are $A \to A - \dd{\theta}$ with the gauge
parameter $\theta$ valued in $\R/2\pi\Z$, invariance of the Wilson line
$W_{\alphae}$ under large gauge transformations requires $\alphae \in \Z$. For
convenience, however, we may rescale electric charges and the gauge field by
$\frac{1}{\qe}$ and magnetic charges by $\qe$ without affecting the above
discussion, ensuring that electric and magnetic charges of dynamical states are
all integers, while lines may carry fractional electric charge. We will take
this convention in the remainder of the paper.

\subsection{Lines in the general case}
In this section we review the main points of \cite{DelZotto:2022ras}, in
particular the notion of a defect group and its relation to the SymTFT. This
generalises the arguments above in three respects: We allow for $r \geq 1$ gauge
fields, we allow dyonic dynamical states as well as lines, and we also allow for
flavour charge. For in-depth discussions of the various charge lattices, see
\cite{Argyres:2022kon,Gaiotto:2010be,Aharony:2013hda}.

We are interested in the 1-form symmetry of a four-dimensional $\U(1)^r$ gauge
theory with charged massive dynamical states. This theory has generalised dyonic
line operators $L$, carrying electric and magnetic charges $\vb*{\alpha} =
(\alphae^1, \alpham^1, \dots, \alphae^r, \alpham^r)$ forming a lattice $\Gamma_L
\subset \R^{2r}$. To be well-defined in the presence of charged states, they
need to satisfy the quantisation condition $\ev{\Gamma_L, \Gamma} \subset \Z$,
where
\begin{equation}
  \ev{\vb*{\alpha},\vb*{\alpha}'}
  = \sum_{i=1}^r \qty(\alphae^i \alpham'^i - \alpham^i \alphae'^i)
\end{equation}
is the Dirac pairing and $\Gamma \subset \Z^{2r}$ is the
lattice of dynamical charges, assumed to be of full rank. We denote this as
$\Gamma_L \subset \Gamma^*$, where $\Gamma^* = \{\vb*{\alpha} \in \R^{2r} \,|\,
\ev{\vb*{\alpha}, \Gamma} \subset \Z\}$ is the dual of $\Gamma$ with respect to
the Dirac pairing.

The worldlines of dynamical states form a subgroup $\Gamma \subset \Gamma_L$. As
these lines can end on a local operator insertion, they cannot be charged under
the one-form symmetry (the local operator is said to \emph{screen} the line).
The 1-form symmetry charges therefore take values in the \emph{defect group}
$\dg = \Gamma^* / \Gamma$. Because the Dirac pairing is perfect, it induces an
isomorphism $\Gamma^* \cong \Hom_\Z(\Gamma, \Z)$ via $\vb*{\alpha} \mapsto
\ev{\cdot, \vb*{\alpha}}$, and we can express the defect group in terms of its
restriction $\QQ\colon \Gamma \to \Hom_\Z(\Gamma, \Z)$ as $\dg \cong
\coker{\QQ}$.

A genuine line operator needs to be well-defined not only in the presence of
dynamical states, but also in the presence of other lines. A configuration of
two lines with charges $\vb*{\alpha}$, $\vb*{\alpha}'$ is subject to a phase
ambiguity $\e^{2\pi\ii\ev{\vb*{\alpha},\vb*{\alpha}'}}$, so this requires the
stronger condition $\ev{\Gamma_L, \Gamma_L} \subset \Z$. There are multiple
maximal solutions $\Gamma_L$ (maximal isotropic, or Lagrangian, sublattices of
$\Gamma^*$) to this constraint, each defining a global structure of the theory.

Let us make this reasoning more concrete. Since the Dirac pairing is
skew-symmetric, there is a basis $\{\gamma^{\mathrm{e},i},
\gamma^{\mathrm{m},i}\}_{i=1}^r$ for $\Gamma$ where it takes the simple form
\begin{align}
  \ev{\gamma^{\mathrm{e},i}, \gamma^{\mathrm{m},j}}
  &= n_i \delta_{ij} \\
  \ev{\gamma^{\mathrm{e},i}, \gamma^{\mathrm{e},j}}
  &= \ev{\gamma^{\mathrm{m},i}, \gamma^{\mathrm{m},j}} = 0
\end{align}
with $n_i \in \N$. Then, since we assume $\Gamma$ has full rank, it is clear
that $\{\frac{1}{n_i}\gamma^{\mathrm{e},i}, \frac{1}{n_i}
\gamma^{\mathrm{m},i}\}_{i=1}^r$ is a basis for $\Gamma^*$, and the defect group
is
\begin{equation}
  \label{eq:defect-group}
  \dg = \Gamma^* / \Gamma \cong \bigoplus_{i=1}^r (\Z / n_i \Z)^2.
\end{equation}
Indeed, the Dirac pairing expressed as a matrix $\QQ^{\alpha\beta} =
\ev{\gamma^\alpha, \gamma^\beta}$ becomes in this basis
\begin{equation}
  \QQ = \bigoplus_{i=1}^r \mqty(0 & n_i \\ -n_i & 0),
\end{equation}
and we see that $\coker{\QQ}$ agrees with \eqref{eq:defect-group}. The integers
$n_i$ are the non-zero invariant factors of $\QQ$---the diagonal entries in its
Smith normal form.

The picture we have outlined is easily generalised to incorporate flavour. We
allow for the possibility of $f$ flavour charges (a rank $f$ flavour group) so
that $\Gamma$ has rank $2r + f$. These have zero Dirac pairing with any other
charge; they generate $\ker{\QQ}$. We quotient out by the flavour charges to
define reduced charge lattices
\begin{equation}
  \tilde{\Gamma} = \frac{\Gamma}{\ker{\QQ}}, \qquad
  \tilde{\Gamma}^* = \frac{\Gamma^*}{\R \otimes_\Z \ker{\QQ}}
\end{equation}
and define the defect group by $\dg = \tilde{\Gamma}^* / \tilde{\Gamma}$. As
above, the Dirac pairing defines an isomorphism $\tilde{\Gamma}^* \cong
\Hom_\Z(\tilde{\Gamma}, \Z)$ with restriction $\tilde{Q}\colon \tilde{\Gamma}
\to \Hom_\Z(\tilde{\Gamma}, \Z)$, such that
\begin{equation}
  \dg \cong \coker{\tilde{\QQ}} \cong \Tor{\coker{\QQ}}
\end{equation}
and
\begin{equation}
  \label{eq:coker-group}
  \coker \QQ \cong \dg \oplus \Z^f.
\end{equation}

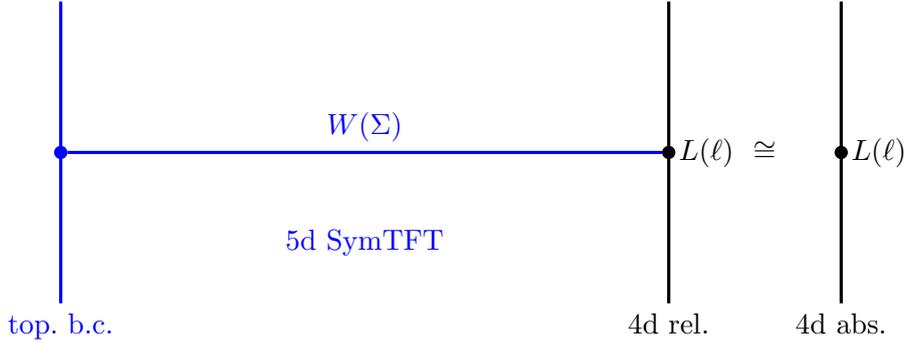
\begin{figure}[t]
  \centering
  \begin{tikzpicture}[scale=4,baseline=(C.base)]
  \draw[very thick] (2,0) -- (2,1);
  \node[below] at (2,0) {4d rel.};
  \node[dotnode] (rel) at (2,0.5) {};
  \node[right] (C) at (rel) {$L(\ell)$};

  \begin{scope}[blue]
    \draw[very thick] (0,0) -- (0,1);
    \node[below] at (0,0) {top.~b.c.};
    \node at (1,0.2) {5d SymTFT};

    \node[dotnode] (top) at (0,0.5) {};
    \draw[very thick] (top) -- (rel) node[midway,above] {$W(\Sigma)$};
  \end{scope}
\end{tikzpicture}
$\cong$
\begin{tikzpicture}[scale=4,baseline=(C.base)]
  \draw[very thick] (0,0) -- (0,1);
  \node[below] at (0,0) {4d abs.};
  \node[dotnode] (abs) at (0,0.5) {};
  \node[right] (C) at (abs) {$L(\ell)$};
\end{tikzpicture}
  \caption{Line operators $L(\ell)$ in the relative 4d theory bound surface
    operators $W(\Sigma)$ in the SymTFT. A line operator is genuine (that is,
    survives in the absolute theory) if the surface operator it bounds can end
    on the topological boundary. See \cite{Bhardwaj:2021mzl} for a detailed
    account.}
  \label{fig:sandwich}
\end{figure}
The central claim of \cite{DelZotto:2022ras} is that the symmetry TFT capturing
the choice of global structure explained above is described by the action
\begin{equation}
  \label{eq:Q-BF}
  S = \frac{\ii}{2\pi}\, \frac{\QQ^{\alpha\beta}}{2}
  \int_{X_5} B_\alpha \wedge \dd{B_\beta}
\end{equation}
where the 4d theory lives on $\M_4 \subset \partial X_5$.\footnote{Importantly,
  $\partial X_5$ can have components other than $\M_4$; in the simplest
  case $X_5 = \M_4 \times [0,1]$ as in \cref{fig:sandwich}.}
The fields $B_\alpha$ are 2-form higher $\U(1)$ gauge fields. (Properly, they
are cocycles in differential cohomology \cite{Cheeger1985,Hopkins:2002rd}; see
also \cite{Freed:2006yc,Hsieh:2020jpj,Apruzzi:2021nmk}. Roughly speaking, they
are locally defined 2-forms such that $\frac{1}{2\pi}\dd{B_\alpha}$ are globally
defined 3-forms with integer periods.\footnote{Compared to the fields
$b_\alpha$ of \cite{DelZotto:2022ras}, our fields are $B_\alpha =
2\pi\,b_\alpha$. This is the conventional normalisation used in physics,
although the $b_\alpha$ are mathematically more natural.}) In the special basis
introduced above, the action reduces to
\begin{equation}
  \label{eq:reduced-BF}
  S = \frac{\ii}{2\pi} \sum_{i = 1}^r n_i \qty(
  \int_{X_5} B_{\mathrm{e},i} \wedge \dd{B_{\mathrm{m},i}}
  - \frac{1}{2} \int_{\partial X_5} B_{\mathrm{e},i} \wedge B_{\mathrm{m},i}).
\end{equation}
We disregard the boundary terms, which are local in the boundary gauge fields,
and focus on the bulk terms, which describe a product of $r$ BF theories
\cite{Witten:1998wy,Kapustin:2014gua}. The terms with $n_i = 1$ describe trivial
(invertible) field theories, but those with $n_i > 1$ give a non-invertible
field theory containing two-dimensional surface operators ending on the 4d line
operators. They form the defect group $\dg$, and their linking relations
\begin{equation}
  \ev{\exp(\ii\oint_\Sigma B_{\mathrm{e},i})
    \exp(\ii\oint_{\Sigma'} B_{\mathrm{m},j})}
  = \exp(\frac{2\pi\ii}{n_i} \delta_{ij} \operatorname{link}(\Sigma, \Sigma'))
\end{equation}
capture the phase ambiguity between 4d lines. A topological boundary condition
determines the set of line operators in the absolute theory; see
\cref{fig:sandwich}. From this perspective it is clear that the defect group is
a property of the SymTFT bulk, while a maximal isotropic sublattice
$\Gamma_L / \Gamma$ is determined by a choice of topological boundary condition,
as derived in the 3d case in \cite{Kapustin:2010hk}.\footnote{The charge lattice
$\Gamma_L / \Gamma$ does not completely specify a topological boundary
condition, but additional data is needed \cite{Lawrie:2023tdz,Gukov:2020btk}.}
In this work, we focus on the SymTFT itself.

\section{Reduction from M-theory}
\label{sec:m-theory}
One derivation of the symmetry TFT of the class $\cS[\su(N), \Sigma_{g,p}]$
theory proceeds by dimensional reduction of M-theory. We realise the $\sN =
(2,0)$ theory on a six-dimensional spacetime $\M_6$ as the worldvolume theory on
a stack of $N$ coincident M5-branes. \Textcite{Witten:1998wy} showed that, in
the limit $N \to \infty$, the near-horizon geometry is $X_7 \times S^4$ where
the conformal boundary of $X_7$ is $\partial X_7 = \M_6$. He further argued that
the low-energy theory close to the branes has a topological sector described by
the Chern--Simons action
\begin{equation}
  \label{eq:witten-CS}
  S = -\frac{\ii N}{2 \cdot 2\pi} \int_{X_7} C \wedge \dd{C},
\end{equation}
where $C$ is the M-theory $\U(1)$ $3$-form gauge field. This can be
justified as follows: Upon reduction of the Chern--Simons term in the Euclidean
11d supergravity action,
\begin{equation}
  S = -\frac{\ii}{6(2\pi)^2}
  \int_{X_7 \times S^4} C \wedge \dd{C} \wedge \dd{C},
\end{equation}
the single-derivative term is
\begin{equation}
  S = -\frac{\ii}{2 \cdot 2\pi} \int_{X_7} C \wedge \dd{C}\
  \frac{1}{2\pi} \oint_{S^4} \dd{C}.
\end{equation}
Since each M5-brane sources one unit of flux for $\dd{C}$, the $S^4$ integral
evaluates to $N$ and we recover \eqref{eq:witten-CS}. This reduction is
well-known, and has been performed in greater generality in the framework of
differential cohomology \cite{Apruzzi:2021nmk}.

Next, we perform Kaluza--Klein reduction on $X_7 = X_5 \times \Sigma_{g,p}$,
beginning with the case $p = 0$, detailed in \cite{Bashmakov:2022uek}. Expand
$C$ in terms of eigen-1-forms of the Laplace--de Rham operator $\Delta =
\dd\delta + \delta\dd$ on $\Sigma_{g,0}$:
\begin{equation}
  C = \sum_i B_i \wedge \omega^i, \qquad
  \Delta \omega^i = \lambda_i \omega^i
\end{equation}
with $\oint_{\Sigma_{g,0}} \omega^i \wedge \hodge \omega^j = 0$ if $\lambda_i
\neq \lambda_j$. The coefficients $B_i$ are 2-forms on $X_5$. Reducing the
kinetic term $\frac{1}{2\kappa^2} \int_{X_7} \dd{C} \wedge \hodge \dd{C}$
produces a mass term for $B_i$ unless $\dd{\omega^i} = 0$; we truncate to these
massless modes. Now reduce the topological term \eqref{eq:witten-CS} to obtain
\begin{equation}
  \label{eq:m-theory-reduced}
  S = -\frac{\ii N}{2 \cdot 2\pi} \sum_{i,j} \int_{X_5} B_i \wedge \dd{B_j}
  \oint_{\Sigma_{g,0}} \omega^i \wedge \omega^j.
\end{equation}
In the terms with $\lambda_i \neq 0$, $\omega^i = \frac{1}{\lambda_i}
\dd{\delta\omega^i}$ is exact and the integral over $\Sigma_{g,0}$ is zero;
similarly if $\lambda_j \neq 0$. Thus the contributing $\omega^i$ are harmonic
forms which we can take to have integral periods. These are by the Hodge theorem
\cite[Theorem~1.45]{Rosenberg1997} and Poincaré duality in bijection with the
generators $c^i$ of $H_1(\Sigma_{g,0}; \Z) \cong \Z^{2g}$. The intersection
pairing $(c^i, c^j) = \int_{\Sigma_{g,0}} \omega^i \wedge \omega^j \in \Z$ is
\begin{align}
  (c^{\mathrm{e},i}, c^{\mathrm{m},j}) &=
  -(c^{\mathrm{m},j}, c^{\mathrm{e},i}) = -\delta^{ij} \\
  (c^{\mathrm{e},i}, c^{\mathrm{e},j}) &=
  (c^{\mathrm{m},i}, c^{\mathrm{m},j}) = 0
\end{align}
with the appropriate choices of generators $c^{\mathrm{e},1},\, \dots,\,
c^{\mathrm{e},g}$, $c^{\mathrm{m},1},\, \dots,\, c^{\mathrm{m},g}$. Thus
\begin{equation}
  \label{eq:m-theory-BF}
  S = \frac{\ii N}{2\pi} \sum_{i = 1}^g
  \int_{X_5} B_{\mathrm{e},i} \wedge \dd{B_{\mathrm{m},i}},
\end{equation}
(again, up to boundary terms) which reproduces \eqref{eq:reduced-BF} with $g$ of
the $n_i$ equal to $N$. By \eqref{eq:defect-group}, this indeed matches with
\eqref{eq:defect-group-claim}. While Witten's construction relied on the
holographic limit $N \to \infty$, we expect the result to hold for any $N$, and
indeed check it for small $N$ in the next section.

The presence of punctures complicates the analysis since there are now boundary
conditions for $C$. Naively, one could observe that since
\eqref{eq:m-theory-reduced} couples the 5d gauge fields through the intersection
pairing on $H_1(\Sigma_{g,0}; \Z)$, and since the elementary cycles surrounding
punctures are in the kernel of the intersection pairing on $H_1(\Sigma_{g,p};
\Z)$, adding punctures should not affect \eqref{eq:m-theory-BF}. The problem is
that by the Hodge decomposition with boundary \cite[Theorem~2.6.1]{Schwarz1995}
and Lefschetz duality \cite[Theorem~3.43]{Hatcher2002}, this pairing results
from the KK reduction above only if $C$ is taken to have Dirichlet boundary
conditions at the punctures, and this is not generally the case. Indeed,
\emph{irregular} punctures do introduce 1-form symmetries ``trapped'' at the
punctures \cite{Bhardwaj:2021mzl}.

From the M-theory perspective, a regular puncture can be described using a
construction of \textcite{Bah:2018jrv,Bah:2019jts}: We modify the space
$\Sigma_{g,0} \times S^4$ in the above construction in a neighbourhood $D
\subset \Sigma_{g,0}$ of the puncture, replacing $D \times S^4$ with a space
$X_6$ whose $\dd{C}$ flux along four-cycles define the puncture data. We thus
obtain the 11-dimensional spacetime $\M_{11} = X_5 \times Y_6$. The space $X_6$
is a fibration of the form $S^2 \to X_6 \to X_4$; $S^1 \to X_4 \to \R^3$ where
the $S^2$ and $S^1$ shrink at certain singular loci.

We outline an argument that this structure means that the puncture does not
modify the defect group. If $X_6$ and $X_4$ were non-singular fibrations, we
would have the long exact sequences
\begin{align}
  \cdots \longto \pi_1(S^2) \longto \pi_1(X_6) \longto \pi_1(X_4)
  \longto \pi_0(S^2) \longto \cdots \\
  \cdots \longto \pi_2(\R^3) \longto \pi_1(S^1) \longto \pi_1(X_4)
  \longto \pi_1(\R^3) \longto \cdots
\end{align}
establishing that $\pi_1(X_6) \cong \pi_1(X_4) \cong \pi_1(S^1) \cong \Z$,
generated by the loop winding along $S^1$. In the present case, however, the
$S^1$ shrinks and this loop is contractible; we therefore expect $\pi_1(X_6) =
0$ instead.\footnote{This is analogous to the singular fibration $S^1 \to S^2
\to [0,1]$ of the 2-sphere over the unit interval. The $S^1$ fibre shrinks to
a point at the endpoints of the interval, so $\pi_1(S^2) = 0$ rather than $\Z$.}
Then $H_1(X_6) = 0$ by the Hurewicz theorem \cite[Theorem~2A.1]{Hatcher2002}.
Now, expanding $C_{\M_{11}} = \sum_i B^{(2)}_{i,X_5} \wedge \omega^{i(1)}_{Y_6}
+ \phi^{(0)}_{X_5} c^{(3)}_{Y_6} + \cdots$, the topological term
\eqref{eq:witten-CS} reduces to
\begin{equation}
  -\frac{\ii}{2\cdot 2\pi} \int_{X_5} B_i \wedge \dd{B_j}
  \oint_{Y_6} \omega^i \wedge \omega^j \wedge \frac{\dd{c}}{2\pi}
\end{equation}
where $\omega^i$ and $\dd{c}/2\pi$ have integral periods; in particular
$\omega^i$ is valued in $H^1(Y_6; \Z)$. Consider for simplicity the case of a
single puncture; then we have $Y_6 = (\Sigma_{g,1} \times S^4) \cup X_6$ with
$(\Sigma_{g,1} \times S^4) \cap X_6 \simeq S^1 \times S^4$, and the reduced
Mayer--Vietoris sequence \cite[section~4.6]{Spanier1966} becomes
\begin{equation}
  \cdots \longto \overbrace{H_1(S^1 \times S^4)}^{\Z} \overset{i_*}{\longto}
  H_1(\Sigma_{g,1} \times S^4) \oplus \overbrace{H_1(X_6)}^0 \longto
  H_1(Y_6) \longto 0
\end{equation}
and it follows that
\begin{equation}
  H_1(Y_6) \cong \frac{H_1(\Sigma_{g,1} \times S^4)}{\im i_*}
  \cong H_1(\Sigma_{g,0} \times S^4) \cong H_1(\Sigma_{g,0}).
\end{equation}
The middle isomorphism above comes from the fact that $i\colon S^1 \times S^4
\to \Sigma_{g,1} \times S^4$ is the inclusion
identifying $S^1$ with the non-contractible\footnote{This loop is
non-contractible whenever $g \geq 1$. The case $g = 0$ corresponds to a sphere
with a single punctures, but we consider only spheres with at least three
punctures.} loop surrounding the puncture; trivialising it amounts to closing
the puncture. The universal coefficient theorem \cite[Theorem~3.2]{Hatcher2002}
now gives $H^1(Y_6; \Z) \cong H^1(\Sigma_{g,0}; \Z)$ and $\omega^i$ are cocycles
on $\Sigma_{g,0}$. Then, the integral of $\frac{\dd{c}}{2\pi}$ captures the
total flux $N$ sourced by the branes; we recover \eqref{eq:m-theory-reduced} and
confirm that the puncture does not affect the defect group.

\section{Defect group from BPS quivers}
\label{sec:bps-quivers}
In this section, we find the defect group and hence the SymTFT using an explicit
construction of the 4d BPS quiver.

\subsection{Full punctures}
\begin{figure}[t]
  \centering
  {
  \newcommand{\firstPart}{
    \begin{scope}[nodes=dotnode]
      \node (p1) at (0, 0) {};
      \node (p2) at (1, 0) {};
      \node (p3) at (60:1) {};
    \end{scope}
    \coordinate (C) at ($1/3*(p1) + 1/3*(p2) + 1/3*(p3)$);
    
    \draw (p1) -- (p2) -- (p3) -- (p1);
  }

  \begin{tikzpicture}[scale=4,baseline=(C.center)]
    \firstPart
    \draw[->] ($ (C) + (-45:.15) $)
          arc[radius=.15,start angle=-45,end angle=225];
  \end{tikzpicture}
  $\longrightarrow$
  \begin{tikzpicture}[scale=4,baseline=(C.center)]
    \begin{scope}[gray]
      \firstPart
      \drawTriangle{3}{p1}{p2}{p3}
    \end{scope}

  \end{tikzpicture}
}
  \caption{Constructing the class $\cS[\su(N)]$ BPS quiver from a triangulation
    ($N = 4$ is shown).}
  \label{fig:SUN-triangle}
\end{figure}
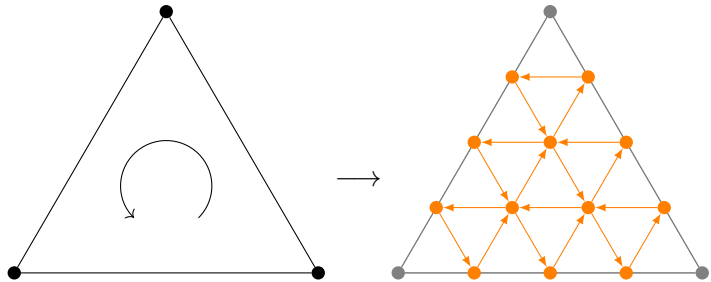
A BPS quiver \cite{Cecotti:2011rv,Alim:2011kw} of a 4d $\sN = 2$
supersymmetric theory has nodes corresponding to charges of certain BPS states,
and arrows such that the signed adjacency matrix is the matrix $\QQ$ of Dirac
pairings as in \cref{sec:review}. The nodes always correspond to physical BPS
states; hence every charge in the lattice they generate is realised by a
dynamical state (not in general a BPS state). We assume, as in
\cite{DelZotto:2022ras}, that this is the full charge lattice of the theory; in
other words, every charge is a sum of BPS charges.

\Textcite{Cecotti:2011rv} described how to construct BPS quivers of class
$\cS[\su(2)]$ theories from ideal triangulations \cite{fomin2007cluster} of
$\Sigma_{g,p}$. The nodes of the quiver are the arcs\footnote{While a
  triangulation and a BPS quiver are both a type of graph, we use distinct
  terminology for them in order to avoid confusion: A triangulation consists of
  \emph{vertices} and \emph{arcs}, while a BPS quiver consists of \emph{nodes}
  and \emph{arrows}.} of the triangulation, and they are joined by arrows going
counterclockwise around each triangle (with respect to the orientation of
$\Sigma_{g,p}$). In the literature on cluster algebras, these are called quivers
of surface type. The procedure generalises to other class $\cS$ theories via
work on Hitchin systems and spectral networks
\cite{Gaiotto:2009hg,Gaiotto:2010be,Gaiotto:2012rg,Gaiotto:2012db} and their
relation to the BPS quivers \cite{Xie:2012dw,Gabella:2017hpz}. We need only the
fact that the BPS quiver for a class $\cS[\su(N)]$ theory with full punctures is
a so-called \emph{$(N - 1)$-triangulation}
\cite{fock2006moduli,Goncharov:2019gzh}: Starting from an ideal triangulation,
the quiver has has $N - 1$ nodes for each arc, as well as $N - 1 \choose 2$
internal nodes in each triangle, connected as in \cref{fig:SUN-triangle}. Once
the quiver is known, the defect group (as well as the flavour rank $f$) can be
extracted using \eqref{eq:coker-group}.

\begin{figure}[t]
  \centering
  {
  \begin{tikzpicture}[scale=4.5,baseline=(C.center)]
    \coordinate (C) at (.5,.45);
    \begin{scope}[nodes=dotnode]
      \node (p1) at (0, 0) {};
      \node (p2) at (1, 0) {};
      \node (p3) at (1, 1) {};
      \node (p4) at (0, 1) {};
    \end{scope}
    \draw (p1) -- (p2) -- (p3) -- (p4) -- (p1);
    \draw (p1) -- (p3);
    \drawTriangle{2}{p1}{p2}{p3}[0]
    \drawTriangle{2}{p1}{p3}{p4}[1]

    \begin{scope}[nodes={below right,blue}]
      \node at (0x1x0) {$0$};
      \node at (0x2x0) {$1$};
      \node at (0x2x1) {$2$};
      \node at (0x1x2) {$3$};
      \node at (1x1x0) {$4$};
      \node at (1x2x0) {$5$};
      \node at (0x1x1) {$6$};
      \node at (1x1x1) {$7$};

      \node at (1x1x2) {$0$};
      \node at (1x2x1) {$1$};
      \node at (1x0x1) {$2$};
      \node at (1x0x2) {$3$};
    \end{scope}

    \draw (.5,-.04) -- (.5,.04);
    \draw (.5,0.96) -- (.5,1.04);
    \draw (-.04,.49) -- (.04,.49);
    \draw (-.04,.51) -- (.04,.51);
    \draw (.96,.49) -- (1.04,.49);
    \draw (.96,.51) -- (1.04,.51);
\end{tikzpicture}
}
  \quad
  \kbalignrighttrue%
  \renewcommand{\kbldelim}{(}%
  \renewcommand{\kbrdelim}{)}%
  $\displaystyle\kbordermatrix{
    & \textcolor{blue}{0} & \textcolor{blue}{1} & \textcolor{blue}{2}
    & \textcolor{blue}{3} & \textcolor{blue}{4} & \textcolor{blue}{5}
    & \textcolor{blue}{6} & \textcolor{blue}{7} \\
    \textcolor{blue}{0} &  0 &  0 &  0 & -1 &  1 &  0 & -1 &  1 \\
    \textcolor{blue}{1} &  0 &  0 & -1 &  0 &  0 &  1 &  1 & -1 \\
    \textcolor{blue}{2} &  0 &  1 &  0 &  0 & -1 &  0 & -1 &  1 \\
    \textcolor{blue}{3} &  1 &  0 &  0 &  0 &  0 & -1 &  1 & -1 \\
    \textcolor{blue}{4} & -1 &  0 &  1 &  0 &  0 &  0 &  1 & -1 \\
    \textcolor{blue}{5} &  0 & -1 &  0 &  1 &  0 &  0 & -1 &  1 \\
    \textcolor{blue}{6} &  1 & -1 &  1 & -1 & -1 &  1 &  0 &  0 \\
    \textcolor{blue}{7} & -1 &  1 & -1 &  1 &  1 & -1 &  0 &  0
  }$
  \caption{BPS quiver and signed adjacency matrix $\QQ$ for the theory
    $\cS[\su(3), \Sigma_{1,1}]$.}
  \label{fig:example}
\end{figure}
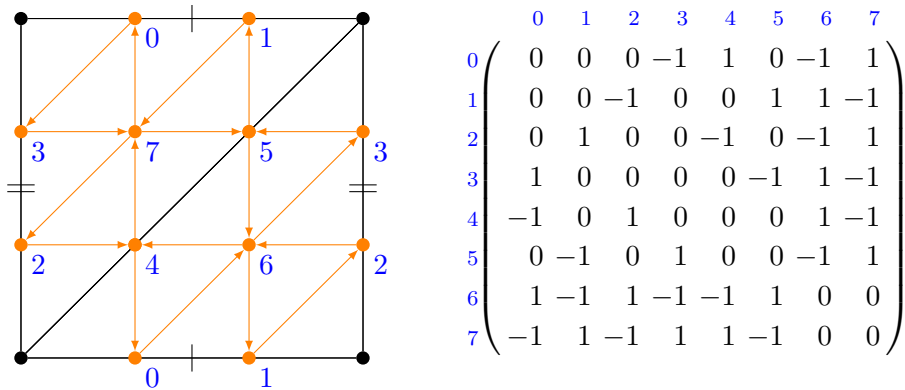
For example, let us put $N = 3$, $g = 1$, $p = 1$ for the class $\cS[\su(3)]$
theory on a torus with one full puncture; see \cref{fig:example}. The torus is
triangulated with two triangles and three arcs that begin and end on the single
puncture. The BPS quiver is the Fock--Goncharov 2-triangulation and has eight
nodes. The corresponding signed adjacency matrix $\QQ$ (\cref{fig:example}) has
the Smith normal form
\begin{equation}
  \operatorname{diag}(1,\, 1,\, 1,\, 1,\, 3,\, 3,\, 0,\, 0)
\end{equation}
so that $\coker{\QQ} \cong (\Z_3)^2 \oplus \Z^2$. This means that the theory has
defect group $\dg \cong (\Z_3)^2$ and two flavour charges.

Accompanying this work is a computer program to carry through this calculation
for arbitrary $N$, $g$ and $p$. Details on the computation are found in
\cref{sec:computation}. We find that
\begin{equation}
  \label{eq:result}
  (\dg \oplus \Z^f)(\cS[\su(N), \Sigma_{g,p}])
  \cong (\Z_N)^{2g} \oplus \Z^{(N - 1)p}
\end{equation}
in agreement with \eqref{eq:defect-group-claim}, for all values $2 \leq N \leq
7$, $0 \leq g \leq 8$ and $1 \leq p \leq 8$ ($3 \leq p \leq 8$ for $g = 0$).

In addition to the defect group, the calculation also determines the rank $r$
and flavour rank $f$ of the theory. \Cref{eq:result} directly gives
\begin{equation}
  \label{eq:flavour-rank}
  f = (N - 1) p.
\end{equation}
To derive $r$, the dimension of the Coulomb branch of the 4d theory, note that
the charge lattice has rank $2r + f$, equal to the number of nodes in the BPS
quiver. As described above, the quiver has $2r + f = {N - 1 \choose 2} t + (N -
1)a$ nodes, where $a$ is the number of arcs in the triangulation and $t$ is the
number of triangles. By Euler's formula $p - a + t = 2 - 2g$, as well as the
fact that $2a = 3t$ in a triangulation, we find that $2r + f = (N^2 - 1)(2g + p
- 2)$. Together with the result \eqref{eq:flavour-rank} for $f$, this leads to
the following formula for the rank:
\begin{equation}
  \label{eq:rank}
  r = (N^2 - 1)(g - 1) + {N \choose 2} p.
\end{equation}

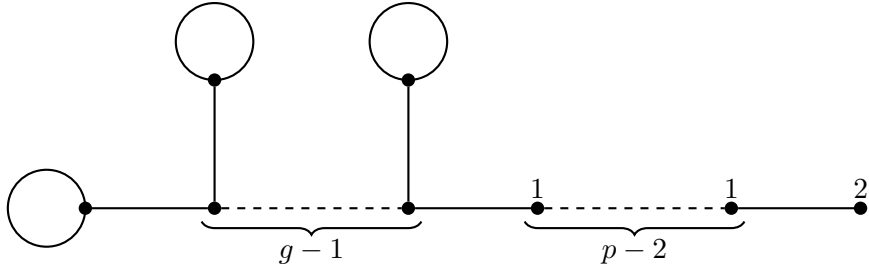
\begin{figure}[t]
  \centering
  {
  \begin{tikzpicture}[scale=1.7,thick]
    \begin{scope}[nodes=dotnode]
      \node (b0) at (0,0) {};
      \node (b1) at (1,0) {};
      \node (b2) at (2.5,0) {};
      \node (f1) at (3.5,0) {};
      \node (f2) at (5,0) {};
      \node (f3) at (6,0) {};
      \node (l1) at (1,1) {};
      \node (l2) at (2.5,1) {};
    \end{scope}

    \draw (b0) -- (b1) -- (l1);
    \draw[dashed] (b1) -- (b2);
    \draw (l2) -- (b2) -- (f1);
    \draw[dashed] (f1) -- (f2);
    \draw (f2) -- (f3);

    \begin{scope}[radius=.3]
      \draw (b0) arc[start angle=0,delta angle=360] (b0);
      \draw (l1) arc[start angle=-90,delta angle=360] (l1);
      \draw (l2) arc[start angle=-90,delta angle=360] (l2);
    \end{scope}
  
    \begin{scope}[decoration={brace,mirror,amplitude=5pt,raise=5pt}]
      \draw[decorate] ($(b1) - (.1,0)$) -- ($(b2) + (.1,0)$)
        node[midway,below,outer sep=8pt] {$g - 1$};
      \draw[decorate] ($(f1) - (.1,0)$) -- ($(f2) + (.1,0)$)
        node[midway,below,outer sep=8pt] {$p - 2$};
    \end{scope}

    \node[above] at (f1) {$1$};
    \node[above] at (f2) {$1$};
    \node[above] at (f3) {$2$};
  \end{tikzpicture}
}
  \caption{Decomposition of $\Sigma_{g,p}$ into $2g + p - 2$ spheres (dots) with
    three full punctures, joined by $3g + p - 3$ tubes (lines). Numbers denote
    punctures.}
  \label{fig:TN}
\end{figure}

We can obtain some basic consistency checks on the above construction and
results constructing $\cS[\su(N), \Sigma_{g,p}]$ by gluing together copies of
the $T_N$ theory \cite{Gaiotto:2009we,Gaiotto:2009gz,Tachikawa:2015bga}, which
is the compactification on a sphere with three full punctures: $T_N =
\cS[\su(N), \Sigma_{0,3}]$. Each full puncture carries an $\SU(N)$ flavour
symmetry \cite{Tachikawa:2013kta} and connecting two punctures by a tube
corresponds to gauging it. Thus we obtain $\Sigma_{g,p}$ from $2g + p - 2$ $T_N$
theories on thrice-punctured spheres after connecting $3g + p - 3$ pairs of
punctures by a tube (\cref{fig:TN}).

While it is not obvious how to extract the defect group from this picture, we
can easily derive $r$ and $f$. The $\SU(N)^p$ flavour symmetry evidently
reproduces \eqref{eq:flavour-rank}. As for the rank, the $T_N$ theory on each
sphere has $d - 2$ Coulomb branch operators of scaling dimension $d$ for each $d
= 3, \dots, N$ \cite[Fact~5.7]{Tachikawa:2015bga}, so its rank is $\sum_{d =
3}^N (d - 2) = {{N - 1} \choose 2}$. Furthermore, each gauged $\SU(N)$
associated to a tube contributes $N - 1$ Coulomb branch operators. The total
rank is
\begin{equation}
  r = {{N - 1} \choose 2} (2g + p - 2) + (N - 1) (3g + p - 3)
\end{equation}
which indeed simplifies to \eqref{eq:rank}.

\subsection{Non-full punctures}
\label{sec:non-full}
\begin{figure}[t]
  \centering
  {
  \newcommand{\rad}{2}
  \newcommand{\n}{5}
  \newcommand{\p}{3}
  \begin{tikzpicture}[scale=2,rotate=45,nodes=dotnode,-latex,baseline=(C.center)]
    \coordinate (C) at (0,0);

    \pgfmathparse{\p-1}
    \foreach \i in {0, ..., \pgfmathresult} {
      \pgfmathparse{360/\p*\i}
      \node (m\i) at (\pgfmathresult:\rad) {};
    }
    \pgfmathparse{\p-1}
    \foreach \i in {0, ..., \pgfmathresult} {
      \pgfmathtruncatemacro{\nextI}{Mod(\i+1,\p)}
      \pgfmathparse{\n-2}
      \foreach \j in {0, ..., \pgfmathresult} {
        \coordinate (mid\i x\j) at ($(m\i)!{(\j+1)/\n}!(m\nextI)$);
        \node (b\i x\j) at ($(mid\i x\j) + ({360/\p*(\i+.5)}:-.3)$) {};
        \node (c\i x\j) at ($(mid\i x\j) + ({360/\p*(\i+.5)}:.3)$) {};
      }
    }
    \pgfmathparse{\p-1}
    \foreach \i in {0, ..., \pgfmathresult} {
      \draw (c\i x0) -- (m\i);
      \draw (m\i) -- (b\i x0);
      \pgfmathtruncatemacro{\prevI}{Mod(\i - 1, \p)}
      \pgfmathtruncatemacro{\prevJ}{\n - 2}
      \draw (c\prevI x\prevJ) -- (m\i);
      \draw (m\i) -- (b\prevI x\prevJ);
      \pgfmathparse{\n-2}
      \foreach \j in {0, ..., \pgfmathresult} {
        \draw[double] (b\i x\j) -- (c\i x\j);
        \pgfmathparse{\j > 0}
        \ifnum\pgfmathresult=1
          \pgfmathtruncatemacro{\prevJ}{\j - 1}
          \draw (c\i x\j) -- (b\i x\prevJ);
          \draw (c\i x\prevJ) -- (b\i x\j);
        \fi
      }
    }
  \end{tikzpicture}
}
  {
  \newcommand{\rad}{2}
  \newcommand{\n}{5}
  \newcommand{\p}{3}
  \begin{tikzpicture}[scale=1,rotate=-15,baseline=(C.center)]
    \coordinate (C) at (0,0);

    \pgfmathparse{\p-1}
    \foreach \i in {0, ..., \pgfmathresult} {
      \pgfmathparse{360/\p*\i}
      \node[circle,draw] (m\i) at (\pgfmathresult:\rad) {$\su(\n)$};
    }
    \pgfmathparse{\p-1}
    \foreach \i in {0, ..., \pgfmathresult} {
      \pgfmathtruncatemacro{\nextI}{Mod(\i+1,\p)}
      \draw (m\i) -- (m\nextI);
    }
  \end{tikzpicture}
}
  \caption{Left: BPS quiver for $\g = \su(N)$ compactified on a torus with
    $p$ simple punctures ($N = 5$, $p = 3$ is shown). \\
    Right: Quiver gauge theory description. A node is a gauge algebra and an
    edge is a bifundamental hypermultiplet.}
  \label{fig:torus-simple-punctures}
\end{figure}
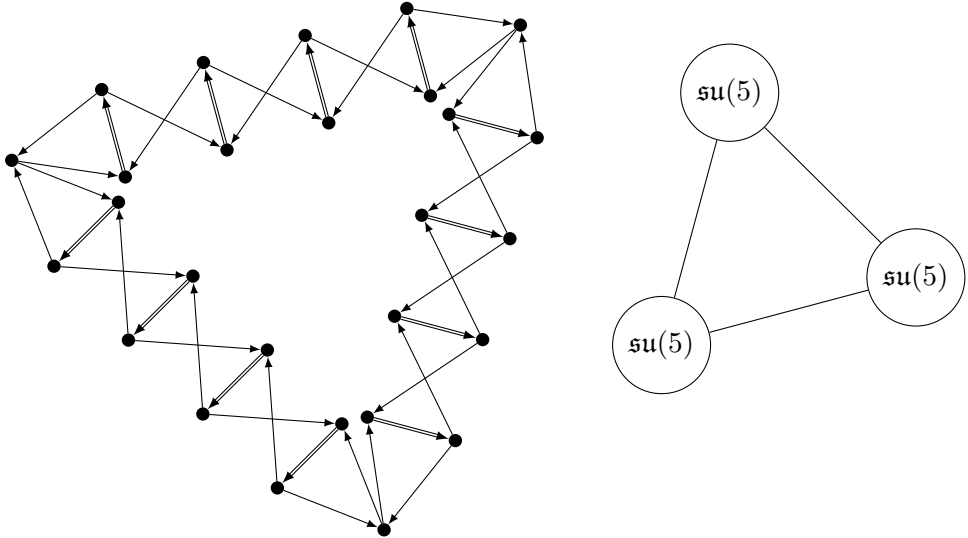
\begin{figure}[t]
  \centering
  {
  \begin{tikzpicture}[scale=1.8,thick]
    \begin{scope}[nodes=dotnode]
      \node (a0) at (0,0) {};
      \node (a1) at (2,0) {};
    \end{scope}

    \draw (a0) -- (a1) node[midway,above] {$SU(N)$};

    \draw[dashed] (a0) -- ($(a0) + (120:.5)$);
    \draw[dashed] (a0) -- ($(a0) + (-120:.5)$);
    \draw[dashed] (a1) -- ($(a1) + (60:.5)$);
    \draw[dashed] (a1) -- ($(a1) + (-60:.5)$);

    \draw[->] (3,0) -- (3.5,0);

    \begin{scope}[nodes=dotnode]
      \node (b0) at (4.5,0) {};
      \node (bm) at (5.5,0) {};
      \node (b1) at (6.5,0) {};
    \end{scope}

    \draw (b0) -- (bm) node[midway,above] (SUN1) {$SU(N)_1$};
    \draw (bm) -- (b1) node[midway,above] (SUN2) {$SU(N)_2$};

    \draw[dashed] (b0) -- ($(b0) + (120:.5)$);
    \draw[dashed] (b0) -- ($(b0) + (-120:.5)$);
    \draw[dashed] (b1) -- ($(b1) + (60:.5)$);
    \draw[dashed] (b1) -- ($(b1) + (-60:.5)$);

    \node[below] at (bm) {\small simple};

    \node (hyp) at (5.5,.7) {\small bifund.~hyper};
    \draw[dotted] (SUN1.north) -- (hyp);
    \draw[dotted] (SUN2.north) -- (hyp);
  \end{tikzpicture}
}
  \caption{Adding a simple puncture.}
  \label{fig:adding-simple}
\end{figure}
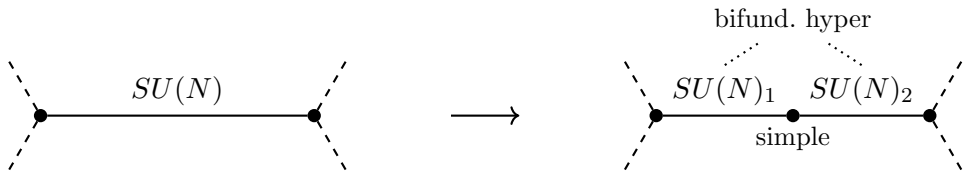
One limitation of our computation is that it deals with full punctures only. In
general one can consider tame punctures labelled by any partition of $N$
\cite{Tachikawa:2013kta}. The general expectation is that the defect group
should be the same for any choice of partitions of $N$ at the punctures, as
argued in \cite{Bhardwaj:2021pfz} and in \cref{sec:m-theory}. Our result
confirms that changing a full puncture, which is labelled by $(1, \dots, 1)$,
into the absence of a puncture, which can be thought of as labelled by $(N)$,
does not affect the defect group. For theories with partial punctures, several
constructions of the BPS quivers exist \cite{Xie:2012dw,Gabella:2017hpz}, but
here we will be content with checking a specific example. Namely, in the case of
$\cS[\su(N), \Sigma_{1,\tilde{p}}]$, where $\Sigma_{1,\tilde{p}}$ denotes the
torus with $p$ \emph{simple} punctures labelled by $(N-1, 1)$, the quiver is
known \cite{Alim:2011kw} and displayed in \cref{fig:torus-simple-punctures}. We
have checked that for this quiver,
\begin{equation}
  \coker{\QQ} \cong (\Z_N)^2 \oplus \Z^p
\end{equation}
for $2 \leq N \leq 7$ and $1 \leq p \leq 8$, so that the defect group is indeed
$(\Z_N)^2$. This Coulomb branch result can also be confirmed from a quiver
gauge theory description (\cref{fig:torus-simple-punctures}) as in
\cite{Bhardwaj:2021pfz}. In a purely electric duality frame, the gauge group is
$\SU(N)^p$ with Wilson lines charged under $Z(\SU(N)^p) \cong (\Z_N^p)^{(1)}$.
The dynamical states in the bifundamental representations screen the 1-form
symmetry to the diagonal $(\Z_N)^{(1)}$, identified with a maximal isotropic
subgroup of $\dg \cong (\Z_N)^2$.

Indeed, this can be slightly generalised to a field-theoretic argument for
arbitrary simple punctures preserving the defect group: Adding a simple puncture
corresponds to replacing a gauged $\SU(N)$ by $\SU(N)_1 \times \SU(N)_2$ and a
hypermultiplet in the bifundamental (\cref{fig:adding-simple}); see
\cite[fig.~12.5]{Tachikawa:2013kta}. The hypermultiplet has $N$-ality $(1, -1)$
and therefore the lines charged under the centre 1-form symmetries
$(\Z_N)^{(1)}_1$ and $(\Z_N)^{(1)}_2$ are identified up to screening.

\begin{figure}[t]
  \centering
  \newcommand{\circlerad}{.6pt}
\newcommand{\ntri}{7}
\newcommand{\myN}{3}
\pgfmathsetmacro{\ltri}{\ntri - 1}
\newcommand{\drawTriangles}{
  \node[dotnode] (O) at (0,0) {};
  \foreach \tri in {0,...,\ltri} {
    \pgfmathparse{\tri * 360/\ntri}
    \node[dotnode] (V\tri) at (\pgfmathresult:1) {};
  }
  \foreach \tri in {0,...,\ltri} {
    \pgfmathparse{int(mod(\tri + 1, \ntri))}
    \drawTriangle{\myN}{O}{V\tri}{V\pgfmathresult}[\tri]
  }
}
\begin{tikzpicture}[scale=3.5,baseline=(O.center)]
  \drawTriangles
  \foreach \tri in {0,...,\ltri} {
    \foreach \n in {\tri x3x0, \tri x2x1, \tri x1x2}
      \draw[very thick,Violet] (\n) circle (\circlerad);
    \foreach \n in {\tri x2x0, \tri x1x1,
      \tri x3x1, \tri x2x2, \tri x1x3,
      \tri x3x0, \tri x2x1, \tri x1x2}
      \draw[blue] (\n) circle (1pt);
  }
\end{tikzpicture}
\begin{tikzpicture}[scale=3.5,baseline=(O.center)]
  \drawTriangles
  \foreach \tri in {0,...,\ltri} {
    \foreach \n in {\tri x3x0, \tri x2x1, \tri x1x2}
      \draw[very thick,Violet] (\n) circle (\circlerad);
    \foreach \n in {\tri x2x0, \tri x1x1}
      \draw[very thick,red] (\n) circle (\circlerad);
    \draw[very thick,Green] (\tri x1x0) circle (\circlerad);
  }
\end{tikzpicture}
  \caption{Left: Construction of a flavour vector $\gamma$ of $\QQ$ around a
    generic puncture, with entries of $1$ at marked nodes and $0$ elsewhere.
    Blue circles mark where contributions from different nodes cancel in
    $\QQ\gamma$. \\
    Right: The $N-1$ such flavour vectors around each puncture. Here $N = 4$ and
    the three flavour vectors are marked in green, red and blue respectively
    (with entries of $1$ at marked nodes and $0$ elsewhere).}
  \label{fig:null-vectors}
\end{figure}
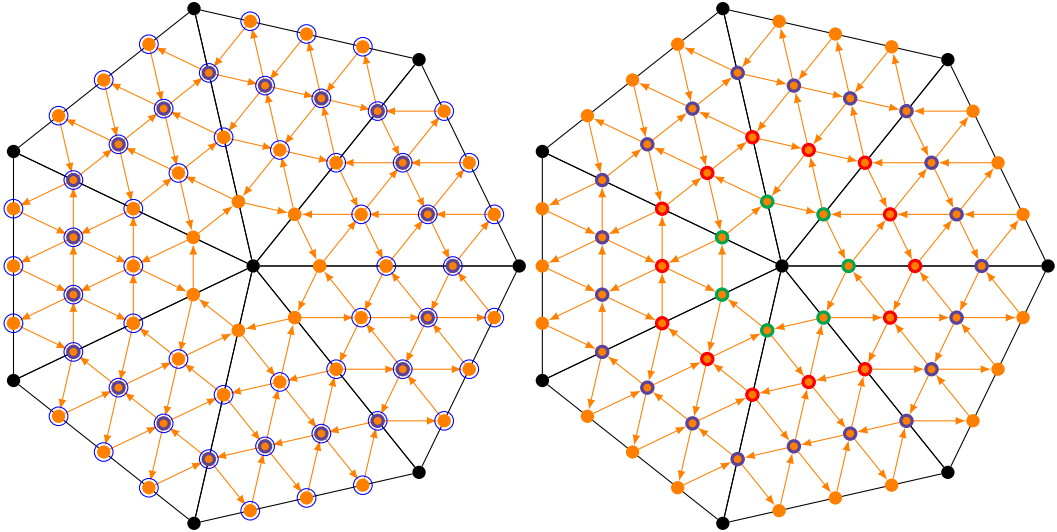
\begin{figure}[t]
  \centering
  \newcommand{\radI}{.6pt}
\newcommand{\radII}{.9pt}
\newcommand{\radIII}{1.2pt}
\newcommand{\drawTrianglesNodes}{
  \node[dotnode] (O) at (0,0) {};
  \node[dotnode] (V0) at (-30:1) {};
  \node[dotnode] (V1) at (30:1) {};
  \node[dotnode] (V2) at (90:1) {};
  \drawNodes{3}{O}{V0}{V1}[0]
  \drawNodes{3}{O}{V1}{V2}[1]
}
\newcommand{\drawTriangleArrows}{
  \drawArrows{3}{O}{V0}{V1}[0]
  \drawArrows{3}{O}{V1}{V2}[1]
}
\begin{tikzpicture}[scale=5,baseline=(O.center)]
  \drawTrianglesNodes
  \draw[red,very thick,->-] (0x2x0) arc[radius=.5,start angle=-30,delta angle=60];
  \draw[red,very thick,->-] (1x2x0) arc[radius=.5,start angle=30,delta angle=60];
  \begin{scope}[Violet,thick]
    \foreach \n in {
      0x1x0, 1x1x0, 1x0x1,
      0x2x0, 0x1x1, 1x2x0, 1x1x1, 1x0x2,
      0x3x0, 0x2x1, 0x1x2, 1x3x0, 1x2x1, 1x1x2, 1x0x3}
      \draw[very thick] (\n) circle (\radI);
    \foreach \n in {0x2x0, 0x1x1, 1x2x0, 1x1x1, 1x0x2,
      0x3x0, 0x2x1, 0x1x2, 1x3x0, 1x2x1, 1x1x2, 1x0x3}
      \draw (\n) circle (\radII);
    \foreach \n in {0x3x0, 0x2x1, 0x1x2, 1x3x0, 1x2x1, 1x1x2, 1x0x3}
      \draw (\n) circle (\radIII);
  \end{scope}
  \drawTriangleArrows
\end{tikzpicture}
\hspace{2em}
\begin{tikzpicture}[scale=5,baseline=(O.center)]
  \drawTrianglesNodes
  \draw[red,very thick,->-] (0x2x0) arc[radius=.5,start angle=-30,delta angle=60];
  \draw[red,very thick,->-] (1x2x0) arc[radius=.5,start angle=-150,delta angle=-60];
  \begin{scope}[Violet,thick]
    \foreach \n in {
      0x1x0, 1x1x0, 1x1x1, 1x1x2, 1x1x3,
      0x2x0, 0x1x1, 1x2x0, 1x2x1, 1x2x2,
      0x3x0, 0x2x1, 0x1x2, 1x3x0, 1x3x1}
      \draw[very thick] (\n) circle (\radI);
    \foreach \n in {
      0x2x0, 0x1x1, 1x2x0, 1x2x1, 1x2x2,
      0x3x0, 0x2x1, 0x1x2, 1x3x0, 1x3x1}
      \draw (\n) circle (\radII);
    \foreach \n in {
      0x3x0, 0x2x1, 0x1x2, 1x3x0, 1x3x1}
      \draw (\n) circle (\radIII);
  \end{scope}
  \drawTriangleArrows
  \node[red] at ($(1x3x0) + (.04,.08)$) {\scriptsize$-4$};
\end{tikzpicture}
  \caption{Charge vectors $\textcolor{Violet}{v}(\textcolor{red}{c})$ mapping to
    torsion elements of $\coker{\QQ}$. Entries of $1$, $2$ and $3$ are marked
    with the corresponding number of blue circles. \\
    Left: a counterclockwise turning path segment, with all zeroes in the image.
    \\
    Right: a direction change, with a single nonzero entry $\pm N$ in the image
    (here $N = 4$).}
  \label{fig:torsion-vectors}
\end{figure}
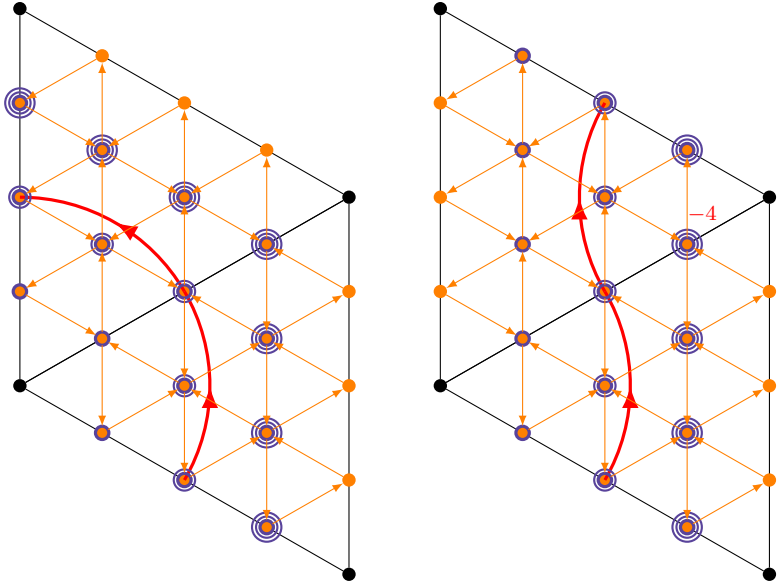
\begin{figure}[t]
  \centering
  \newcommand{\scl}{3}
\newcommand{\radI}{.8pt}
\newcommand{\radII}{1.3pt}
\newcommand{\radIII}{1.8pt}
\newcommand{\drawTrianglesNodes}{
  \node (C) at (0,-.035) {};
  \node[dotnode] (O) at (0,0) {};
  \node[dotnode] (V0) at (-30:1) {};
  \node[dotnode] (V1) at (30:1) {};
  \drawNodes{3}{O}{V0}{V1}[0]
}
\newcommand{\drawTriangleArrows}{
  \drawArrows{3}{O}{V0}{V1}[0]
}

\begin{equation*}
  \Huge
  \begin{tikzpicture}[scale=\scl,baseline=(C.center)]
    \drawTrianglesNodes
    \draw[red,very thick,->-] (0x2x0)
      arc[radius=.5,start angle=150,delta angle=-60];
    \drawTriangleArrows
    \begin{scope}[Violet,thick]
      \foreach \n in {
        0x1x0, 0x1x1, 0x1x2, 0x1x3,
        0x2x0, 0x2x1, 0x2x2,
        0x3x0, 0x3x1}
      \draw[very thick] (\n) circle (\radI);
      \foreach \n in {
        0x2x0, 0x2x1, 0x2x2,
        0x3x0, 0x3x1}
      \draw (\n) circle (\radII);
      \foreach \n in {0x3x0, 0x3x1}
      \draw (\n) circle (\radIII);
    \end{scope}
  \end{tikzpicture}
  \hspace{.4ex} +
  \begin{tikzpicture}[scale=\scl,baseline=(C.center)]
    \drawTrianglesNodes
    \draw[red,very thick,->-] (0x2x2)
      arc[radius=.5,start angle=-90,delta angle=-60];
    \drawTriangleArrows
    \begin{scope}[Violet,thick]
      \foreach \n in {
        0x0x1, 0x1x1, 0x2x1, 0x3x1,
        0x0x2, 0x1x2, 0x2x2,
        0x0x3, 0x1x3}
      \draw[very thick] (\n) circle (\radI);
      \foreach \n in {
        0x0x2, 0x1x2, 0x2x2,
        0x0x3, 0x1x3}
      \draw (\n) circle (\radII);
      \foreach \n in {0x0x3, 0x1x3}
      \draw (\n) circle (\radIII);
    \end{scope}
  \end{tikzpicture}
  \hspace{.4ex} =
  \begin{tikzpicture}[scale=\scl,baseline=(C.center)]
    \drawTrianglesNodes
    \draw[red,very thick,->-] (0x2x0)
      arc[radius=.5,start angle=-30,delta angle=60];
    \drawTriangleArrows
    \begin{scope}[Violet,thick]
      \foreach \n in {
        0x1x0, 0x0x1,
        0x2x0, 0x1x1, 0x0x2,
        0x3x0, 0x2x1, 0x1x2, 0x0x3}
      \draw[very thick] (\n) circle (\radI);
      \foreach \n in {
        0x2x0, 0x1x1, 0x0x2,
        0x3x0, 0x2x1, 0x1x2, 0x0x3}
      \draw (\n) circle (\radII);
      \foreach \n in {0x3x0, 0x2x1, 0x1x2, 0x0x3}
      \draw (\n) circle (\radIII);
    \end{scope}
  \end{tikzpicture}
  \hspace{.4ex} + 4
  \begin{tikzpicture}[scale=\scl,baseline=(C.center)]
    \drawTrianglesNodes
    \drawTriangleArrows
    \begin{scope}[Violet,thick]
      \foreach \n in {0x3x1, 0x2x2, 0x1x3}
      \draw[very thick] (\n) circle (\radI);
    \end{scope}
  \end{tikzpicture}
\end{equation*}
\begin{equation*}
  \Huge
  \begin{tikzpicture}[scale=\scl,baseline=(C.center)]
    \drawTrianglesNodes
    \draw[red,very thick,->-] (0x0x2)
      arc[radius=.5,start angle=30,delta angle=-60];
    \drawTriangleArrows
    \begin{scope}[Violet,thick]
      \foreach \n in {
        0x3x0, 0x2x1, 0x1x2, 0x0x3,
        0x2x0, 0x1x1, 0x0x2,
        0x1x0, 0x0x1}
        \draw[very thick] (\n) circle (\radI);
      \foreach \n in {
        0x2x0, 0x1x1, 0x0x2,
        0x1x0, 0x0x1}
        \draw (\n) circle (\radII);
      \foreach \n in {0x1x0, 0x0x1}
        \draw (\n) circle (\radIII);
    \end{scope}
  \end{tikzpicture}
  \hspace{.4ex} = -
  \begin{tikzpicture}[scale=\scl,baseline=(C.center)]
    \drawTrianglesNodes
    \draw[red,very thick,->-] (0x2x0)
    arc[radius=.5,start angle=-30,delta angle=60];
    \drawTriangleArrows
    \begin{scope}[Violet,thick]
      \foreach \n in {
        0x1x0, 0x0x1,
        0x2x0, 0x1x1, 0x0x2,
        0x3x0, 0x2x1, 0x1x2, 0x0x3}
      \draw[very thick] (\n) circle (\radI);
      \foreach \n in {
        0x2x0, 0x1x1, 0x0x2,
        0x3x0, 0x2x1, 0x1x2, 0x0x3}
      \draw (\n) circle (\radII);
      \foreach \n in {0x3x0, 0x2x1, 0x1x2, 0x0x3}
      \draw (\n) circle (\radIII);
    \end{scope}
  \end{tikzpicture}
  \hspace{.4ex} + 4
  \begin{tikzpicture}[scale=\scl,baseline=(C.center)]
    \drawTrianglesNodes
    \drawTriangleArrows
    \begin{scope}[Violet,thick]
      \foreach \n in {
        0x1x0, 0x0x1,
        0x2x0, 0x1x1, 0x0x2,
        0x3x0, 0x2x1, 0x1x2, 0x0x3}
      \draw[very thick] (\n) circle (\radI);
    \end{scope}
  \end{tikzpicture}
\end{equation*}
  \caption{The assignment $v$ of charges to cycles is additive modulo $N$ (here
    $N = 4$). Entries of $1$, $2$ and $3$ are marked with the corresponding
    number of blue circles.}
  \label{fig:homomorphism}
\end{figure}
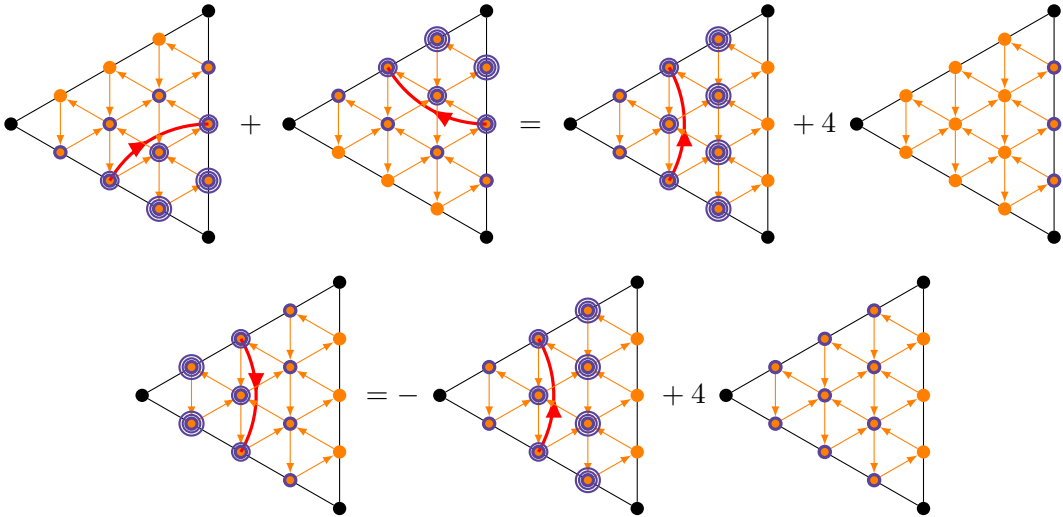

\subsection{Structure of the BPS quiver}
\label{sec:structure}
The result \eqref{eq:result} is a purely combinatorial statement about the
$(N-1)$-triangulations of \textcite{fock2006moduli}, and it is interesting to
consider it as such. It has been partially addressed in the mathematical
literature; in particular, \cref{eq:flavour-rank} when $N = 2$ is a special case
of Theorem~14.3 of \cite{fomin2007cluster}. Here, we present some
observations on the structure of the quivers that make the result plausible in
the form of a proof sketch. It would be interesting to see if this reasoning
could be extended to a full elementary proof of \eqref{eq:result}.

Recall that we have the charge lattice $\Gamma = \Z^{2r + f}$ with standard
basis given by the nodes of the quiver, and we conceptualise the Dirac pairing
as a $\Z$-linear map $\QQ\colon \Gamma \to \Hom_\Z(\Gamma, \Z)$.

First, the free factor in $\coker{\QQ}$ being $\Z^{(N-1)p}$ is equivalent to
$\ker{\QQ} \cong \Z^{(N-1)p}$. It is in fact easy to exhibit $(N-1)p$ null
(flavour) vectors; $N-1$ for each elementary cycle wrapping a puncture in
$\Sigma_{g,p}$, as in \cref{fig:null-vectors}. Recall that a node in the quiver
is a generator of $\Gamma$; in the figure, the marked nodes $\gamma_i$ define a
vector $\gamma = \sum_i \gamma_i \in \Gamma$. The covector
$\QQ\gamma = \ev{\cdot,\gamma} \in \Hom_\Z(\Gamma,\Z)$ is found by following the
arrows adjacent to each marked point, adding $1$ for outgoing arrows and $-1$
for incoming arrows. The construction of $\gamma$ ensures that all such
contributions (occurring at the nodes marked with blue circles) cancel, so that
$\QQ\gamma = 0$. This works no matter the number of arcs incident to the
puncture. This shows that the rank of $\ker{\QQ}$ is at least $(N-1)p$, but does
not rule out hypothetical further flavour vectors.\footnote{The story is
somewhat complicated by torsion. The cycles form a maximal linearly independent
subset of $\ker{\QQ}$, but not a generating set. For example, consider the case
where the vertices of each triangle are all distinct. Summing all cycles gives a
multiple of the null vector $\gamma = (1, \dots, 1)$, namely $3\gamma$, but
$\gamma$ itself cannot in general be attained this way.}

Next we search for the torsional part of $\coker{\QQ}$; the defect group. There
is an explicit map $v\colon H_1(\Sigma_{g,p}; \Z) \to \Gamma$, defined in
\cref{fig:torsion-vectors}, such that
$\im(\QQ \circ v) \subset N \Hom_\Z(\Gamma, \Z)$.
\Cref{fig:homomorphism} shows that it defines a homomorphism $v_N\colon
H_1(\Sigma_{g,p}; \Z_N) \to \Gamma \otimes \Z_N$.
Each generator $c \in H_1(\Sigma_{g,p}; \Z)$ passes through a sequence of
triangles, and in each triangle, turns either clockwise or counterclockwise. A
cycle that turns purely clockwise or counterclockwise surrounds a single
puncture; then $v(c)$ is a sum of flavour vectors as in \cref{fig:null-vectors}.
Thus $v_N$ descends to a homomorphism $\tilde{v}_N\colon H_1(\Sigma_{g,0}; \Z_N)
\to \tilde{\Gamma} \otimes \Z_N$ such that $(\tilde{\QQ} \otimes 1_{\Z_N}) \circ
\tilde{v}_N = 0$ (recall that $\tilde{\Gamma} = \Gamma / \ker{\QQ}$,
$\tilde{\Gamma}^* \cong \Hom_\Z(\tilde{\Gamma}, \Z)$ and $\dg =
\coker(\tilde{\QQ}\colon \tilde{\Gamma} \hookrightarrow \tilde{\Gamma}^*)$).

Here we conjecture that in fact $\ker{(\tilde{\QQ} \otimes 1_{\Z_N})} =
\im{\tilde{v}_N}$, so that we have an exact sequence
\begin{equation}
  \label{eq:ZN-Tor-LES}
  0 \longto H_1(\Sigma_{g,0}; \Z_N) \xrightarrow{\tilde{v}_N}
  \tilde{\Gamma} \otimes \Z_N \xrightarrow{\tilde{Q} \otimes 1_{\Z_N}}
  \tilde{\Gamma}^* \otimes \Z_N \longto
  \dg \otimes \Z_N \longto 0.
\end{equation}
In other words, $\Tor(\dg, \Z_N) \cong H_1(\Sigma_{g,0}; \Z_N) \cong
(\Z_N)^{2g}$. From this, if one can show that $\dg$ is $N$-periodic ($N\dg =
0$, that is, $N\tilde{\Gamma}^* \subset \im{\tilde{\QQ}}$); then $\dg \cong
\Tor(\dg, \Z_N)$ and \eqref{eq:result} follows.

\section*{Acknowledgements}
This work is supported by the European Research Council under grant 851931
(MEMO). I thank Michele Del Zotto for suggesting the project and for his
mentorship throughout. I am grateful to Azeem Hasan, Robert Moscrop and Shani
Nadir Meynet for enlightening discussions. I thank the Department of Physics of
the University of Colorado, Boulder and the Simons Collaboration on Global
Categorical Symmetries for hospitality during the TASI 2023 school and the
GCS2023 conference and school, respectively. Finally, I am grateful to the
anonymous referee, whose comments greatly helped improve the paper's clarity.

\appendix
\section{Computation}
\label{sec:computation}
The computations in \cref{sec:bps-quivers} were done using SageMath
\cite{sagemath}. The source code and results are available at
\url{https://gitlab.com/eliasrg/class-s-defect-groups}.

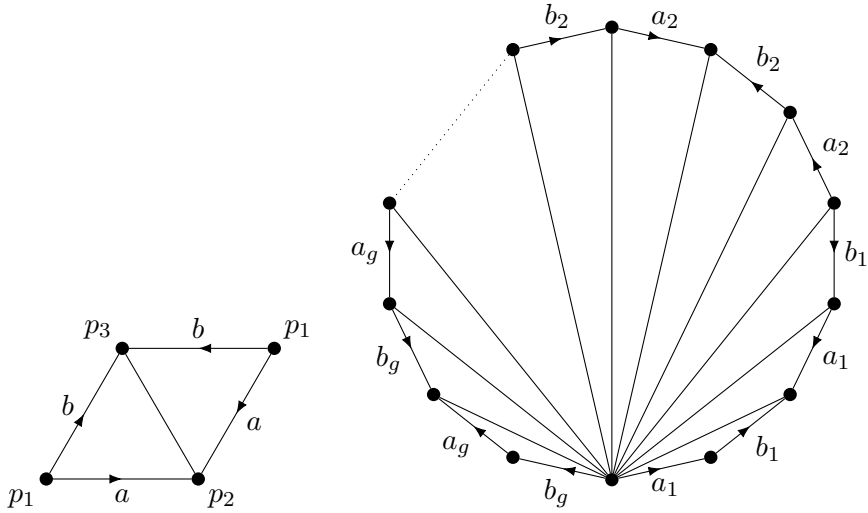
\begin{figure}[t]
  \centering
  \begin{tikzpicture}[scale=2]
  \begin{scope}[nodes=dotnode]
    \node (p1) at (0, 0) {};
    \node (p2) at (1, 0) {};
    \node (p3) at (60:1) {};
    \node (p4) at ($ (p2) + (p3) $) {};
  \end{scope}

  \node[below left] at (p1) {$p_1$};
  \node[below right] at (p2) {$p_2$};
  \node[above left] at (p3) {$p_3$};
  \node[above right] at (p4) {$p_1$};
  
  \begin{scope}[nodes={midway,auto}]
    \draw[->-] (p1) -- (p2) node[swap] {$a$};
    \draw[->-] (p4) -- (p2) node {$a$};
    \draw[->-] (p1) -- (p3) node {$b$};
    \draw[->-] (p4) -- (p3) node[swap] {$b$};
    \draw (p2) -- (p3);
  \end{scope}
\end{tikzpicture}
  \begin{tikzpicture}[scale=3]
  \foreach \i in {0, ..., 13} {
    \unless\ifnum\i=9
      \node[dotnode] (p\i) at ({-90 + 360/14*\i}:1) {};
    \fi
  }

  \begin{scope}[nodes={midway, auto}]
    \draw[->-] (p0) -- node[swap] {$a_1$} (p1);
    \draw[->-] (p1) -- node[swap] {$b_1$} (p2);
    \draw[->-] (p3) -- node {$a_1$} (p2);
    \draw[->-] (p4) -- node {$b_1$} (p3);

    \draw[->-] (p4) -- node[swap] {$a_2$} (p5);
    \draw[->-] (p5) -- node[swap] {$b_2$} (p6);
    \draw[->-] (p7) -- node {$a_2$} (p6);
    \draw[->-] (p8) -- node {$b_2$} (p7);

    \draw[->-] (p10) -- node[swap] {$a_g$} (p11);
    \draw[->-] (p11) -- node[swap] {$b_g$} (p12);
    \draw[->-] (p13) -- node {$a_g$} (p12);
    \draw[->-] (p0) -- node {$b_g$} (p13);

    \draw[dotted] (p8) -- (p10);

    \foreach \i in {2, 3, ..., 8, 10, 11, 12}
      \draw[thin] (p0) -- (p\i);
  \end{scope}
\end{tikzpicture}
  \caption{Ideal triangulations of $\Sigma_{0,3}$ and $\Sigma_{g,1}$
    for $g > 0$.}
  \label{fig:starting-triangulations}
\end{figure}

\begin{figure}[t]
  \centering
  {
  \newcommand{\firstPart}{
    \begin{scope}[nodes=dotnode]
      \node (p1) at (0, 0) {};
      \node (p2) at (1, 0) {};
      \node (p3) at (60:1) {};
    \end{scope}
    \coordinate (C) at ($1/3*(p1) + 1/3*(p2) + 1/3*(p3)$);
    
    \draw (p1) -- (p2) -- (p3) -- (p1);
  }

  \begin{tikzpicture}[scale=3,baseline=(C.center)]
    \firstPart
  \end{tikzpicture}
  $\longrightarrow$
  \begin{tikzpicture}[scale=3,baseline=(C.center)]
    \firstPart
    \node[dotnode] (p4) at (C) {};
    \draw (p1) -- (p4);
    \draw (p2) -- (p4);
    \draw (p3) -- (p4);
  \end{tikzpicture}
}
  \caption{Adding a puncture in the interior of a triangle.}
  \label{fig:adding-puncture}
\end{figure}
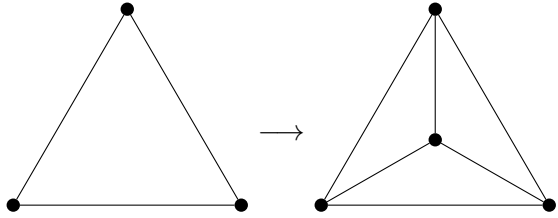

The first step is to construct an ideal triangulation of the punctured Riemann
surface $\Sigma_{g,p}$. We achieve this by starting from a triangulation of
$\Sigma_{g,1}$, or of $\Sigma_{0,3}$ if $g = 0$. These are shown in
\cref{fig:starting-triangulations}. We then add additional punctures one by one,
as shown in \cref{fig:adding-puncture}, until there are $p$ punctures in total.
In preparation for the next step, we also keep track of the orientation of each
triangle.

Having obtained the triangulation, we next construct the class $\cS[\su(N)]$ BPS
quiver. We put $N - 1$ nodes on each arc and $N - 1 \choose 2$ internal nodes in
each triangle, and connect them with arrows according to
\cref{fig:SUN-triangle}, counterclockwise with respect to the orientation of the
surface, as in \cite[fig.~23]{Goncharov:2019gzh}.

Finally, we compute the defect group and flavour rank from the quiver's
signed adjacency matrix $\QQ$ using \eqref{eq:coker-group} and
\begin{equation}
  \coker \QQ \cong \coker S
\end{equation}
where $S$ is the Smith normal form of $\QQ$. The result of this computation is
\cref{eq:result}.

\section{Mathematical preliminaries}
\label{sec:math}
In this appendix, we collect the most important standard mathematical theorems
used in the main text for the reader's convenience.

\paragraph{The skew normal form}
The existence of the special basis of the charge lattice $\Gamma$ follows from
\cite[Theorem~IV.1]{Newman1972}: If $\QQ$ is a skew-symmetric $n \times n$
matrix with integer entries, then $\QQ$ is congruent to a block-diagonal matrix
$\QQ'$ (meaning that there exists an integer matrix $B$, invertible over the
integers, such that $\QQ = B^T \QQ' B$) of the form
\begin{equation}
  \QQ' = \bigoplus_{i=1}^r \mqty(0 & n_i \\ -n_i & 0) \oplus 0_{f \times f}
\end{equation}
where $n = 2r + f$ for some $r$ and $f$ (in particular, $\Omega$ has even rank
$2r$, and the kernel is isomorphic to $\Z^f$).

\paragraph{Hodge isomorphism for manifolds with boundary}
For a manifold $X$ without boundary, the Hodge theorem
\cite[Theorem~1.45]{Rosenberg1997} states that there is an isomorphism between
the group of harmonic $p$-forms (the kernel of the Laplacian $\Delta = \dd\delta
+ \delta\dd$ on $p$-forms) and the $p$-th de Rham cohomology of $X$. In other
words, every class in $H^p(X;\R) \cong H_\mathrm{dR}^p(X)$ has a unique harmonic
representative. There is a version of this for manifolds with boundary
\cite[Theorem~2.6.1, Corollary~2.6.2]{Schwarz1995}, which asserts isomorphisms
\begin{equation}
  H^p(X;\R) \cong \mathcal{H}^p_N(X) \qq{and}
  H^p(X,\partial X;\R) \cong \mathcal{H}^p_D(X)
\end{equation}
where $\mathcal{H}^p_N(X)$ and $\mathcal{H}^p_D(X)$ are the harmonic $p$-forms
with Neumann and Dirichlet boundary conditions on $\partial X$, respectively.
This holds when $X$ is a compact so-called $\partial$-manifold (orientable
smooth manifold with boundary that is complete as a metric space). \\

\noindent The remaining facts are about algebraic topology. Most of them are
found, with more detailed statements, in \textcite{Hatcher2002}.

\paragraph{Fibrations}
A \emph{fibration} in the usual sense is a map $p\colon E \to B$ such that any
homotopy $h\colon X \times [0,1] \to B$ can be lifted along any map
$g\colon X \to E$ such that $p \circ g = h(\mbox{--}, 0)$; that is, there exists
a map $\tilde{h}\colon X \times [0,1] \to E$ with $p \circ \tilde{h} = h$. The
fibres $p^{-1}(b)$ are all homotopy equivalent to some $F$
\cite[Theorem~4.61]{Hatcher2002}, and one writes $F \to E \to B$. The homotopy
groups fit into a long exact sequence \cite[Theorem~4.41]{Hatcher2002}:
\begin{equation}
  \cdots \longto \pi_n(F) \longto \pi_n(E) \longto \pi_n(B) \longto
  \pi_{n-1}(F) \longto \cdots.
\end{equation}
The singular fibrations discussed in \cref{sec:m-theory} are not of this type;
in particular, above certain singular points, the fibre shrinks to a point and
not all fibres are homotopy equivalent.

\paragraph{The Hurewicz theorem for $H_1$}
The Hurewicz theorem \cite[Theorems~2A.1, 4.32]{Hatcher2002} describes the
relationship between the first nontrivial homotopy and homology groups of a
space. For the argument in \cref{sec:m-theory}, we have made use of the
connection between $\pi_1$ and $H_1$, namely that for a path-connected space
$X$, the first homology group $H_1(X)$ is isomorphic to the abelianisation of
the fundamental group $\pi_1(X)$. In particular, if $\pi_1(X) = 0$, then $H_1(X)
= 0$.

\paragraph{The reduced Mayer--Vietoris sequence}
The \emph{reduced} homology groups \cite[section~4.3]{Spanier1966}
$\tilde{H}_n(X)$ of a space (or chain complex) $X$ are the same as $H_n(X)$ for
$n \geq 1$, but $H_0(X) \cong \tilde{H}_0(X) \oplus \Z$. In particular, they
have the convenient property that $\tilde{H}_0(X) = 0$ if $X$ is path-connected.

For a space expressed as the union $A \cup B$ of two subspaces, the
\emph{reduced Mayer--Vietoris sequence} \cite[section~4.6]{Spanier1966} is an
exact sequence
\begin{equation}
  \cdots \longto \tilde{H}_n(A \cap B) \overset{i_*}{\longto}
  \tilde{H}_n(A) \oplus \tilde{H}_n(B) \overset{j_*}{\longto}
  \tilde{H}_n(A \cup B) \overset{\partial_*}{\longto}
  \tilde{H}_{n-1}(A \cap B) \longto \cdots.
\end{equation}
Here $i_* = i_{A*} \oplus (-i_{B*})$ and $j_* = j_{A*} + j_{B*}$, where
$A \cap B \xhookrightarrow{i_A} A \xhookrightarrow{j_A} A \cup B$ and
$A \cap B \xhookrightarrow{i_B} B \xhookrightarrow{j_B} A \cup B$ are the
inclusions.

\paragraph{The universal coefficient theorem for cohomology}
For a space $X$, the universal coefficient theorem for cohomology
\cite[Theorem~3.2]{Hatcher2002} asserts the existence of split short exact
sequences
\begin{equation}
  0 \longto \Ext(H_{n-1}(X), G) \longto H^n(X;G)
  \longto \Hom(H_n(X), G) \longto 0
\end{equation}
for any $n$ and any coefficient group $G$. Here $\Ext(H, G)$ is functorial in
$H$ and $G$ and in particular $\Ext(H, G) = 0$ if $H$ is a free abelian group.
Therefore, since $H_0(X)$ is free, there is an isomorphism
$H^1(X;G) \cong \Hom(H_1(X), G)$.

\paragraph{Lefschetz duality}
Poincaré duality \cite[Theorem~3.30]{Hatcher2002} states that, for a closed
orientable $d$-dimensional manifold $X$, there is an isomorphism $H^n(X) \cong
H_{d-n}(X)$. Lefschetz duality \cite[Theorem~3.43]{Hatcher2002} is the analogous
statement for manifolds with boundary: If $X$ is a compact orientable
$d$-dimensional manifold with boundary, then there are isomorphisms
$H^n(X,\partial X) \cong H_{d-n}(X)$ and $H^n(X) \cong H_{d-n}(X,\partial X)$.

\paragraph{The $\Tor$ functor}
The $\Tor$ functor was used to interpret the exact sequence
\eqref{eq:ZN-Tor-LES}. Given an abelian group $G$, the functor $G \otimes
\mbox{--}\colon \mathbf{Ab} \to \mathbf{Ab}$ is right exact but not in general
left exact. Its first left derived functor is called $\Tor$ (see
e.g.~\cite[section~2.4]{GelfandManin1999}). Concretely, this means the
following. Every abelian group $H$ is a quotient of a free abelian group, that
is, there is an exact sequence $0 \to F_1 \to F_0 \to H \to 0$ with $F_0, F_1$
free (a free resolution of $H$). Then the sequence $G \otimes F_1 \to G \otimes
F_0 \to G \otimes H \to 0$ is exact, but the first morphism is not in general
injective. Its kernel is known as $\Tor(G, H)$, so that the extended sequence
\begin{equation}
  0 \longto \Tor(G, H) \longto G \otimes F_1 \longto G \otimes F_0
  \longto G \otimes H \longto 0
\end{equation}
is exact. $\Tor(G, H)$ is functorial in $G$ and $H$ and satisfies in particular
\cite[Proposition~3A.5]{Hatcher2002} that $\Tor(\Z_N, H) \cong
\{h \in H\,|\,Nh = 0\}$, the subgroup of $N$-torsion elements of $H$.

\printbibliography

\end{document}